\def\year{2018}\relax
\documentclass[letterpaper]{article} 
\usepackage{aaai18}  
\usepackage{times}  
\usepackage{helvet}  
\usepackage{courier}  
\usepackage{url}  
\usepackage{graphicx}  
\usepackage{amsfonts}
\usepackage{amsmath}
\usepackage{color}
\usepackage{tabularx}
\usepackage{multirow}
\usepackage{paralist}

\begin{document}
	%
	\title{Faithful to the Original: Fact Aware Neural Abstractive Summarization}
	\author{
			Ziqiang Cao$^{1,2}$\thanks{Contribution during internship at Microsoft Research}   ~~ ~~ Furu Wei$^3$ ~~ ~~ Wenjie Li$^{1,2}$ ~~ ~~ Sujian Li$^4$\\
		$^1$Department of Computing, The Hong Kong Polytechnic University, Hong Kong\\
		$^2$Hong Kong Polytechnic University Shenzhen Research Institute, China \\
		$^3$Microsoft Research, Beijing, China\\
		$^4$Key Laboratory of Computational Linguistics, Peking University, MOE, China \\
		{\tt \{cszqcao, cswjli\}@comp.polyu.edu.hk  } \\
		{\tt furu@microsoft.com} \\
		{\tt lisujian@pku.edu.cn} \\
	}
	\maketitle
	\begin{abstract}
		Unlike extractive summarization, abstractive summarization has to fuse different parts of the source text, which inclines to create fake facts.
		Our preliminary study reveals nearly 30\% of the outputs from a state-of-the-art neural summarization system suffer from this problem.
		While previous abstractive summarization approaches usually focus on the improvement of informativeness, we argue that faithfulness is also a vital prerequisite for a practical abstractive summarization system.
		To avoid generating fake facts in a summary, we leverage open information extraction and dependency parse technologies to extract actual fact descriptions from the source text.
		The dual-attention sequence-to-sequence framework is then proposed to force the generation conditioned on both the source text and the extracted fact descriptions.
		Experiments on the Gigaword benchmark dataset demonstrate that our model can greatly reduce fake summaries by 80\%.
		Notably, the fact descriptions also bring significant improvement on informativeness since they often condense the meaning of the source text.
	\end{abstract}
	
	\section{Introduction}
	The exponentially growing online information has necessitated the development of effective automatic summarization systems.
In this paper, we focus on an increasingly intriguing task, i.e., abstractive sentence summarization~\cite{rush2015neural} which generates a shorter version of a given sentence while attempting to preserve its original meaning. 
This task is different from document-level summarization since it is hard to apply the common extractive techniques~\cite{over2001introduction}.
Selecting existing sentences to form the sentence summary is impossible.
Early studies on sentence summarization involve handcrafted rules~\cite{zajic2007multi}, syntactic tree pruning~\cite{knight2002summarization} and statistical machine translation techniques~\cite{banko2000headline}. 
Recently, the application of the attentional sequence-to-sequence (s2s) framework has attracted growing attention in this area \cite{rush2015neural,chopra2016abstractive,nallapati2016abstractive}.

As we know, sentence summarization inevitably needs to fuse different parts in the source sentence and is abstractive.
Consequently, the generated summaries often mismatch with the original relations and yield fake facts. 
Our preliminary study reveals that nearly 30\% of the outputs from a state-of-the-art s2s system suffer from this problem.
Previous researches are usually devoted to increasing summary informativeness. 
However, one of the most essential prerequisites for a practical abstractive summarization system is that the generated summaries must accord with the facts expressed in the source.
We refer to this aspect as summary faithfulness in this paper.
A fake summary may greatly misguide the comprehension of the original text.
Look at an illustrative example of the generation result using the state-of-the-art s2s model \cite{nallapati2016abstractive} in Table~\ref{tb:invalid_example}.
The actual subject of the verb ``postponed'' is ``repatriation''.
Nevertheless, probably because the entity ``bosnian moslems'' is closer to ``postponed'' in the source sentence, the summarization system wrongly regards ``bosnian moslems'' as the subject and counterfeits a fact ``bosnian moslems postponed''.
Meanwhile, the s2s system generates another fake fact: ``unhcr pulled out of bosnia'' and puts it into the summary.
Consequently, although the informativeness (ROUGE-1 F1=0.57) and readability of this summary are high, its meaning departs far from the original.
This sort of summaries is nearly useless in practice.
	
	\begin{table}
		\centering
		\begin{tabularx}{\linewidth}{p{1cm}|X}
			\hline
			Source & the repatriation of at least \#,\#\#\# bosnian moslems was postponed friday after the unhcr pulled out of the first joint scheme to return refugees to their homes in northwest bosnia . \\ \hline
			Target & \textbf{repatriation} of bosnian moslems postponed                                                                                       \\ \hline
			s2s & \textbf{bosnian moslems} postponed after unhcr pulled out of \textbf{bosnia}                                                                                  \\ \hline
		\end{tabularx}
		\caption{An example of fake summaries generated by the state-of-the-art s2s model. ``\#'' stands for a digit masked during preprocessing.}
		\label{tb:invalid_example}
	\end{table}
	
	Since the fact fabrication is a serious problem, intuitively, encoding existing facts into the summarization system should be an ideal solution to avoid fake generation.
	To achieve this goal, the first step is to extract the facts from the source sentence.
	In the relatively mature task of Open Information Extraction (OpenIE)~\cite{banko2007open}, a fact is usually represented by a relation triple consisting of (subject; predicate; object).
	For example, given the source sentence in Table~\ref{tb:invalid_example}, the popular OpenIE tool~\cite{angeli2015leveraging} generates two relation triples including \textit{(repatriation; was postponed; friday)} and \textit{(unhcr; pulled out of; first joint scheme)}.
	Obviously, these triples can help rectify the mistakes made by the s2s model.
	However, the relation triples are not always extractable, e.g., from the imperative sentences.
Hence, we further adopt a dependency parser and supplement with the (subject; predicate) and (predicate; object) tuples identified from the parse tree of the sentence.  
This is also inspired by the work of parse tree based sentence compression (e.g., \cite{knight2002summarization}).
We represent a fact through merging words in a triple or tuples to form a short sentence, defined as a \textit{fact description}.
	Fact descriptions actually form the skeletons of sentences.
	Thus we incorporate them as an additional input source text in our model.
	Our experiments reveal that the words in the extracted fact descriptions are 40\% more likely to be included in the actual summaries than the entire words in the source sentences.
	That is, fact descriptions clearly provide the right guidance for summarization.
	Next, using both source sentence and fact descriptions as input, we extend the state-of-the-art attentional s2s model~\cite{nallapati2016abstractive} to fully leverage their information.
	Specially, we use two Recurrent Neural Network (RNN) encoders to read the sentence and fact descriptions in parallel.
	With respective attention mechanisms, our model computes the sentence and fact context vectors.
	It then merges the two vectors according to their relative reliabilities.
	Finally, a RNN decoder makes use of the integrated context to generate the summary word-by-word.
	Since our summarization system encodes \textbf{facts} to enhance \textbf{faithfulness}, we call it \textbf{FTSum}.
	
	To verify the effectiveness of FTSum, we conduct extensive experiments on the Gigaword sentence summarization benchmark dataset~\cite{rush-chopra-weston:2015:EMNLP}.
	The results show that our model greatly reduces the fake summaries by 80\% compared to the state-of-the-art s2s framework.
	Due to the compression nature of fact descriptions, the use of them also brings the significant improvement in terms of automatic informativeness evaluation. 
	The contributions of our work can be summarized as follows:
	\begin{itemize}
		\item To the best of our knowledge, we are the first to explore the faithfulness problem of abstractive summarization.
		\item We propose a dual-attention s2s model to push the generation to follow the original facts.
		\item Since the fact descriptions often condense the meaning of
		the source sentence, they also bring the significant benefit to promote informativeness.
	\end{itemize}

	\section{Fact Description Extraction}
	Based on our observation, 30\% of summaries generated by state-of-the-art s2s models suffer from fact fabrication, such as the mismatch between the predicate and its subject or object.
	Therefore, we propose to explicitly encode existing fact descriptions into the model.
	We leverage popular tools of Open Information Extraction (OpenIE) and dependency parser for this purpose.
	OpenIE refers to the extraction of entity relations from the open-domain text. 
	In OpenIE, a fact is typically interpreted as a relation triple consisting of (subject; predicate; object).
	We join all the items in a triple (i.e., subject + predicate + object) since it usually acts as a concise sentence.
	An example of the OpenIE outputs is presented in Table~\ref{tb:openie_triple}.
	As we can see,  OpenIE may extract multiple triples to reflect an identical fact in different granularities.
	In some extreme cases, one relation can yield over 50 triple variants, which brings high redundancy and burdens the computation cost of the model.
	To balance redundancy and fact completeness, we remove a relation triple if all its words are covered by another one.
	For example, only the last fact description (i.e., I saw cat sitting on desk) in Table~\ref{tb:openie_triple} is reserved.
	When different fact descriptions are extracted at the end, we use a special separator ``$|||$'' to concatenate them to accelerate the encoding process, which is explained by Eq.~\ref{eq:0_1} and \ref{eq:reset}.
	
	\begin{table}[t]
		\centering
		\begin{tabular}{l|l}
			\hline
			Sentence                   & I saw a cat sitting on the desk \\ \hline
			\multirow{3}{*}{Triples} & (I; saw; cat)                \\ \cline{2-2} 
			& (I; saw; cat sitting)           \\ \cline{2-2} 
			& (I; saw; cat sitting on desk)   \\ \hline
		\end{tabular}
		\caption{Examples of OpenIE triples in different granularities. We extract the following fact description: \textit{I saw cat sitting on desk}}
		\label{tb:openie_triple}
	\end{table}
	
	\begin{figure*}
		\centering
		\includegraphics[width=0.75\linewidth]{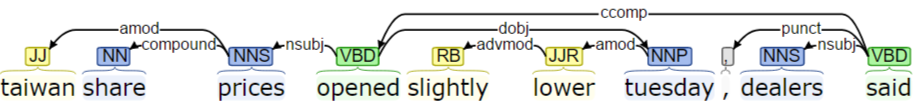}
		\caption{A dependency tree example. The meaning of the dependency labels can be referred to \cite{de2008stanford}. We extract the following two fact descriptions: \textit{taiwan share prices opened lower tuesday} $|||$ \textit{dealers said}}
		\label{fig:depparse}
	\end{figure*}
	
	OpenIE is able to give a complete description of the entity relations.
	However, it is worth noting that, the relation triples are not always extractable, e.g., from the imperative sentences.
	In fact, about 15\% of the OpenIE outputs are empty on our dataset.
	These empty instances are likely to damage the robustness of our model.
	As observed, although the complete relation triples are not always available, the (subject; predicate) or (predicate; object) tuples are almost present in each sentence.
	Therefore, we leverage the dependency parser to dig out the appropriate tuples to supplement the fact descriptions.
	A dependency parser converts a sentence into the labeled (governor; dependent) tuples.
	We extract the predicate-related tuples according to the labels: \textit{nsubj}, \textit{nsubjpass}, \textit{csubj}, \textit{csubjpass} and \textit{dobj}.
	To acquire more complete fact descriptions, we also reserve the important modifiers including the adjectival (\textit{amod}), numeric (\textit{nummod}) and noun compound (\textit{compound}).
	We then merge the tuples containing the same words, and order words based on the original sentence to form the fact descriptions.
	Take the dependency tree in Fig.~\ref{fig:depparse} as an example.
	The output of OpenIE is empty for this sentence.
	Based on the dependency parser, we firstly filter the following predicate-related tuples: \textit{(prices; opened) (opened; tuesday) (dealers; said)} and the modify-head tuples: \textit{(taiwan; price) (share; price)  (lower; tuesday)}.
	These tuples are then merged to form two fact descriptions: \textit{taiwan share prices opened lower tuesday} $|||$ \textit{dealers said}.  
	
	In the experiments, we employ the popular NLP pipeline Stanford CoreNLP~\cite{manning-EtAl:2014:P14-5} to handle OpenIE and dependency parse at the same time. 
	We combine the fact descriptions derived from both parts, and screen out the fact descriptions with the pattern ``somebody said/declared/announced'', which are usually meaningless and insignificant.
	Referring to the copy ratios in Table~\ref{tb:relation_info}, words in fact descriptions are 40\% more likely to be used in the summary than the words in the original sentence.
	It indicates that fact descriptions truly condense the meaning of sentences to a large extent.
	The above statistics also supports the practice of dependency parse based compressive summarization~\cite{knight2002summarization}.
	However, the length sum of extracted fact descriptions is shorter than the actual summary in 20\% of the sentences, and 4\% of the sentences even hold empty fact descriptions.
	In addition, from Table~\ref{tb:relation_info} we can find that on average one key source word is missing in the fact descriptions.
	Thus, without the source sentence, we cannot reply on fact descriptions alone to generate summaries. 
	
	\begin{table}[ht]
		\centering
		\begin{tabular}{l|ll}
			\hline
			Source:  & Sentence & Fact \\ \hline
			AvgLen   & 31.4     & 18.2      \\
			Count & 1        & 2.7       \\
			Copy\%  & 0.12      & 0.17         \\ \hline
		\end{tabular}
		\caption{Comparisons between source sentences and relations. AvgLen is the average number of tokens. Copy\% means the proportion of source tokens can be found in the summary.}
		\label{tb:relation_info}
	\end{table}
	
	\section{Fact Aware Neural Summarization}

	\subsection{Model Framework}
\begin{figure*}
		\centering
		\includegraphics[width=0.99\linewidth]{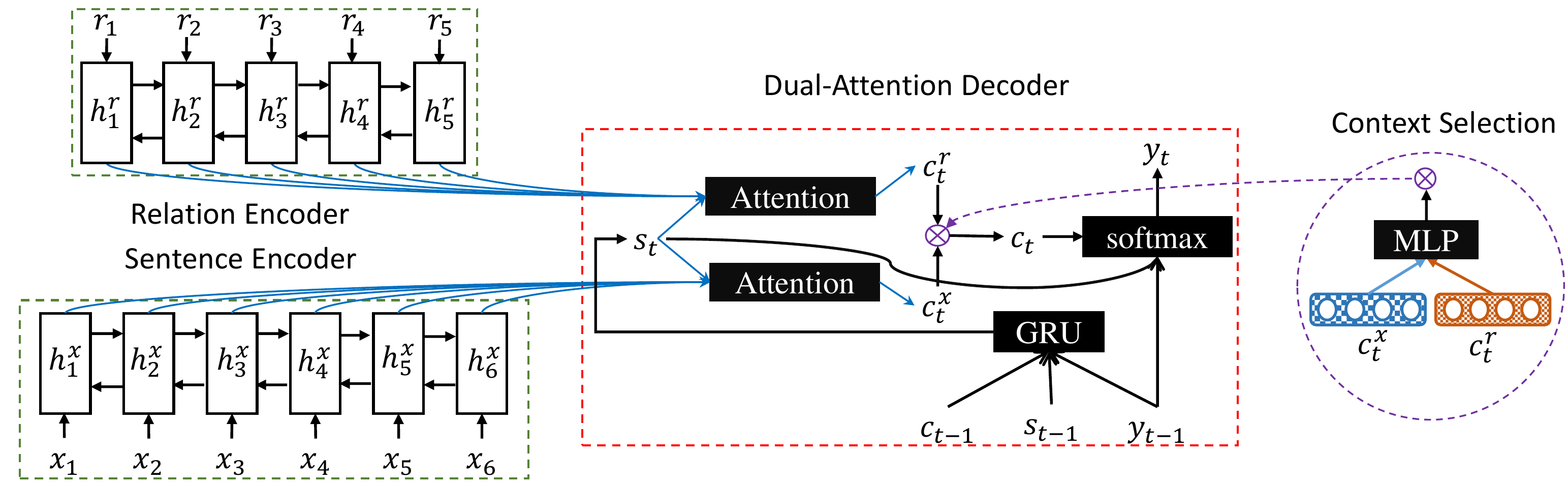}
		\caption{Model framework}
		\label{fig:modelframework}
	\end{figure*}

	As shown in Figure~\ref{fig:modelframework}, our model consists of three modules including two encoders and a dual-attention decoder equipped with a context selection gate network.
	The sentence encoder reads the input words
	${\mathbf{x}} = ({x_1}, \cdots {x_n})$ and builds its corresponding representation $({\mathbf{h}_1^x}, \cdots \mathbf{h}_n^x)$. 
	Likewise, the relation encoder converts the fact descriptions ${\mathbf{r}} = ({r_1}, \cdots {r_k})$ into hidden states $({\mathbf{h}_1^r}, \cdots \mathbf{h}_k^r)$.
	With the respective attention mechanisms, our model computes the sentence and relation context vectors ($\mathbf{c}_t^x$ and $\mathbf{c}_t^r$) at each decoding time step $t$.
	The gate network is followed to merge the context vectors according to their relative associations with the current generation.
	The decoder produces summaries ${\mathbf{y}} = ({y_1}, \cdots {y_l})$ word-by-word conditioned on the tailored context vector which embeds the semantics of both source sentence and fact descriptions.

	\subsection{Encoders}
	The input includes the source sentence $\mathbf{x}$ and the fact descriptions $\mathbf{r}$.
	For each sequence, we employ the bidirectional Gated Recurrent Unit (BiGRU) encoder~\cite{cho2014learning}, to construct its semantic representation.
	Take the sentence $\mathbf{x}$ as an example.
	The GRU at the time step $i$ is defined as follows:
	\begin{equation}\label{eq:GRU}
		{{\bf{h}}_i} = {\text{GRU}}({x_i},{{\bf{h}}_{i - 1}})
	\end{equation}
	The BiGRU consists of a forward GRU and a backward GRU. 
	Suppose the corresponding outputs are $({{\mathbf{\overset{\lower0.5em\hbox{$\smash{\scriptscriptstyle\rightarrow}$}}{h} }}_1}, \cdots {{\mathbf{\overset{\lower0.5em\hbox{$\smash{\scriptscriptstyle\rightarrow}$}}{h} }}_n})$ and $({{\mathbf{\overset{\lower0.5em\hbox{$\smash{\scriptscriptstyle\leftarrow}$}}{h} }}_1}, \cdots {{\mathbf{\overset{\lower0.5em\hbox{$\smash{\scriptscriptstyle\leftarrow}$}}{h} }}_n})$, respectively.
	Then, the composite hidden state of a word is the concatenation of the two GRU representations, i.e., ${{\mathbf{h}}_i} = [{{\mathbf{\overset{\lower0.5em\hbox{$\smash{\scriptscriptstyle\rightarrow}$}}{h} }}_i};{{\mathbf{\overset{\lower0.5em\hbox{$\smash{\scriptscriptstyle\leftarrow}$}}{h} }}_i}]$.
	
	For the relation sequence $\mathbf{r}$, since it contains multiple independent fact descriptions, we introduce boundary indicators $\mathbf{\gamma}$ to separate their hidden states.
	Specially, the value of $\mathbf{\gamma}$ is defined as follows:
	\begin{equation} \label{eq:0_1}
		{\gamma _i} = \left\{ {\begin{array}{*{20}{l}}
				{0,{r_i} \text{ is ``$|||$''}}\\
				{1,\text{otherwise}}
		\end{array}} \right.
	\end{equation}
	Then, ${\mathbf{\gamma}}$ is used to reset the GRU state in Eq.~\ref{eq:GRU}:
	\begin{equation} \label{eq:reset}
		{\bf{h'}}_i = {\gamma _i}{{\bf{h}}_i}
	\end{equation}
	In this way, all the fact descriptions will start with the same zero vector.
	In other words, they are encoded \textit{independently}.
	Finally, both sentence hidden states $\{ {\bf{h}}_i^x\} $ and relation hidden states $\{ {\bf{h}}_i^r\} $ are fed to the decoder. 
	
	\subsection{Dual-Attention Decoder}
	Previous s2s models have developed some task-specific modifications on the decoder, such as to incorporate  the copying mechanism~\cite{gu2016incorporating} and coverage mechanism~\cite{see2017get}.
	As this paper focuses on the faithfulness problem, we use the most popular decoder, i.e., GRU with attentions~\cite{bahdanau2014neural}.
	At each decoding time step $t$, GRU reads the previous output $y_{t - 1}$ and context vector ${{\mathbf{c}}_{t - 1}}$ as inputs to compute new hidden state $\mathbf{s}_t$:
	\begin{equation}
		{{\bf{s}}_t} = {\rm{GRU}}({y_{t - 1}},{{\bf{c}}_t},{{\bf{s}}_{t - 1}})
	\end{equation}
	Since we have both sentence and relation representations as input, we develop two attentional layers to construct the overall context vector ${{\mathbf{c}}_{t}}$. 
	For instance, the context representation of the sentence at time step $t$ is computed as \cite{luong2015effective}:
	\begin{align}
		{e_{t,i}^x} &= \text{MLP}({{\bf{s}}_{t}},{{\bf{h}}_i^x})\\
		{\alpha _{t,i}^x} &= \frac{{\exp ({e_{t,i}^x})}}{{\sum\nolimits_j {\exp ({e_{t,j}^x})} }} \\
		{{\bf{c}}_t^x} &= \sum\nolimits_i {{\alpha _{t,i}^x}{{\bf{h}}_i^x}}, 
	\end{align}
	where MLP stands for multi-layer perceptrons.
	The context vector of the relation ${\bf{c}}^r$ can be computed similarly.
	We combine ${\bf{c}}_t^x$ and ${\bf{c}}_t^r$ to build the overall context vector ${\bf{c}}_t$.
	We explore two alternative combination approaches.
	The first one is called ``FTSum$_c$'', which simply concatenates two context vectors: 
	\begin{equation}\label{eq:context_concate}
		{{\bf{c}}_t} = [{\bf{c}}_t^x;{\bf{c}}_t^r]
	\end{equation}
	The other approach is denoted as ``FTSum$_g$'', where we also use MLP to build a gate network and combine context vectors with the weighted sum:
	\begin{align}
		{{\bf{g}}_t} &= {\rm{MLP}}({\bf{c}}_t^x,{\bf{c}}_t^r) \label{eq:context_gate}\\
		{{\bf{c}}_t} &= {{\bf{g}}_t} \odot {\bf{c}}_t^x + ({\bf{1}} - {{\bf{g}}_t}) \odot {\bf{c}}_t^r,
	\end{align}
	where ``$\odot$'' means the element-wise dot.
	Experiments show that FTSum$_g$ significantly outperforms FTSum$_c$, and the gate values apparently reflect the relative reliability of sentence and fact descriptions. 
	
	Finally, the softmax layer is introduced to generate the next word based on previous word $y_{t-1}$, context vector ${\bf{c}}_t$ and current decoder state $\mathbf{s}_t$.
	\begin{align}
		{{\bf{o}}_t} &= {{\bf{W}}_w}[{y_{t - 1}}] + {{\bf{W}}_c}{{\bf{c}}_t} + {{\bf{W}}_s}{{\bf{s}}_t} \\
		p({y_t}|{y_{<t}}) &= {\rm{softmax}}({{\bf{W}}_o}{{\bf{o}}_t})
	\end{align}
	where $\mathbf{W}_.$ stands for a weight matrix.
	
	\subsection{Learning}
	The learning goal is to maximize the estimated probability of the actual summary. 
	We adopt the common negative log-likelihood (NLL) as the loss function.
	\begin{equation}
		J(\mathbf{\theta} ) =  - \frac{1}{{|{\mathbf{D}}|}}\sum\limits_{({\mathbf{x}},{\mathbf{r}},{\mathbf{y}}) \in {\mathbf{D}}} {\log (p({\mathbf{y}}|{\mathbf{x}},{\mathbf{r}})} ),
	\end{equation}
	where $\mathbf{D}$ denotes the training dataset and $\mathbf{\theta}$ stands for the model parameters.
	We use Adam~\cite{kingma2014adam} with mini-batches as the optimization algorithm. 
	We set the learning rate $\alpha=0.001$ and the mini-batch size to 32.
	Similar to \cite{zhou2017selective}, we evaluate the model performance on the development set for every 2000 batches and halve the learning rate if the cost increases for 10 consecutive validations. 
	In addition, we apply gradient clipping \cite{pascanu2013difficulty} with range $[-5,5]$ during training to enhance the stability of the model.

	\section{Experiments}
	\subsection{Datasets}
	We conduct experiments on the Annotated English Gigaword corpus, as with \cite{rush-chopra-weston:2015:EMNLP}.
	This parallel corpus is produced by pairing the first sentence in the news article and its headline as the summary with heuristic rules.
	The training and development datasets are built through the script\footnote{\url{https://github.com/facebook/NAMAS}} released
	by \cite{rush-chopra-weston:2015:EMNLP}. 
	The script also performs various basic text normalization, including tokenization, lower-casing, replacing all digit characters with \#, and mask the words appearing less than 5 times with a UNK tag. 
	It comes up with about 3.8M sentence-headline pairs as the training set and 189K pairs as the development set.
	We use the same Gigaword test set as \cite{rush-chopra-weston:2015:EMNLP}.
	It contains 2000 sentence-headline pairs.
	Following \cite{rush2015neural}, we remove pairs with empty titles, leading to slightly different accuracy compared with \cite{rush-chopra-weston:2015:EMNLP}. 
	The statistics of the Gigaword corpus is presented in Table~\ref{tb:dataset}.
	
	\begin{table}[]
		\centering
		\begin{tabular}{l|lll}
			\hline
			Dataset      & Train & Dev. & Test \\ \hline
			Count        & 3.8M  & 189k & 1951 \\
			AvgSourceLen & 31.4  & 31.7 & 29.7     \\
			AvgTargetLen & 8.3   & 8.3  & 8.8     \\ \hline
		\end{tabular}
		\caption{Data statistics for the English Gigaword. AvgSourceLen is the average input sentence length and AvgTargetLen is the average headline length.}
		\label{tb:dataset}
	\end{table}

	\subsection{Evaluation Metric}
	We adopt ROUGE~\cite{lin2004rouge} for automatic evaluation.
	ROUGE has been the standard evaluation metric for DUC shared tasks since 2004. 
	It measures the quality of summary by computing overlapping lexical units between the candidate summary and actual summaries, such as unigram, bigram and longest common subsequence (LCS). 
	Following the common practice, we report ROUGE-1 (unigram), ROUGE-2 (bi-gram) and ROUGE-L (LCS) F1 scores\footnote{We use the ROUGE evaluation option: -m -n 2 -w 1.2} in the following experiments.
	ROUGE-1 and ROUGE-2 mainly consider informativeness while ROUGE-L is supposed to be linked to readability.
	
	In addition, we manually inspect whether the generated summaries accord with the facts in the original sentences.
	We mark summaries into three categories: FAITHFUL, FAKE and UNCLEAR.
	The last one refers to the case where a generated summary is too incomplete to judge its faithfulness, such as just producing a UNK tag.
	
	\subsection{Implementation Details}
	Since the dataset has already masked infrequent words with the UNK tag, we reserve all the rest words in the training set.
	As a result, the sizes of source and target vocabularies are 120k and 69k, respectively. 
	With reference to \cite{nallapati2016abstractive}, we leverage the popular s2s framework dl4mt\footnote{\url{https://github.com/kyunghyuncho/dl4mt-material}} as the starting point, and set the size of word embeddings to 200.
	We initialize word embeddings with GloVe~\cite{pennington2014glove}.
	All the GRU hidden state dimensions are fixed to 400.
	We use dropout~\cite{srivastava2014dropout} with probability $p = 0.5$.
	With the decoder, we use the beam search of size 6 to generate the summary, and restrict the maximal length of a summary to 20 words.
	We find that the average system summary length from all our models (about 8.0 words) is very much consistent with that of the ground truth on the development set, without any special tuning.

	\subsection{Baselines}
	We compare our proposed model with the following six state-of-the-art baselines:
	\begin{description}
		\item[ABS] \cite{rush2015neural} used an attentive CNN encoder and NNLM decoder to summarize the sentence.
		\item[ABS+] \cite{rush2015neural}
		further tuned the ABS model with additional features to balance
		the abstractive and extractive tendency.
		\item[RAS-Elman] As the extension of the ABS model, it used a convolutional attention-based encoder and an RNN decoder \cite{chopra2016abstractive}.
		\item[Feats2s]  \cite{nallapati2016abstractive} used a full
		s2s RNN model and added the hand-crafted features such as POS tag and NER, to enhance the encoder representation.
		\item[Luong-NMT] \cite{luong2015effective} applied the two-layer LSTMs Neural machine translation model with 500 hidden
		units in each layer.
		\item[att-s2s] We implement the standard attentional s2s with dl4mt, and denote this baseline as ``att-s2s''.
	\end{description}
	
	\subsection{Informativeness Evaluation}
	At first, look at the final cost values during training in Table~\ref{tb:ppl}.
	We can see that our model achieves the lowest perplexity compared against the state-of-the-art systems.
	It is also noted that, FTSum$_g$ largely outperforms FTSum$_c$, which verifies the importance of context selection.
	The ROUGE F1 scores are then reported in Table~\ref{tb:rouge}.
	Although the focus of our model focuses is to improve faithfulness, the ROUGE scores it receives are also much higher than the other methods.
	Note that, ABS+ and Feats2s have utilized a series of hand-crafted features, but our model is totally data-driven.
	Even though, our model surpasses Feats2s by 13\% and ABS+ by 56\% on ROUGE-2.
	When fact descriptions are ignored, our model is equivalent to the standard attentional s2s model s2s+att.
	Therefore, it is safe to conclude that, fact descriptions have significant contribute to the increase of ROUGE scores.
	One probable reason is that fact descriptions are much more informative than the original sentence, as shown in Table~\ref{tb:relation_info}.
	It also largely explains why FTSum$_g$ is superior to FTSum$_c$.
	FTSum$_c$ treats the source sentence and relations equally, while FTSum$_g$ tells the fact descriptions are often more reliable, as discussed in more detail later.
	
	\begin{table}
		\centering
		\begin{tabular}{l|l}
			\hline
			Model     & Perplexity \\ \hline
			ABS$^\dag$       & 27.1       \\
			RAS-Elman$^\dag$ & 18.9       \\
			s2s-att   & 24.5       \\
			FTSum$_c$     & 20.1       \\
			FTSum$_g$     & \textbf{16.4}       \\ \hline
		\end{tabular}
		\caption{Final perplexity on the development set. $^\dag$ indicates the value is cited from the corresponding paper. ABS+, Feats2s and Luong-NMT do not provide this value.}
		\label{tb:ppl}
	\end{table}
	
	\begin{table}
		\centering
		\begin{tabular}{l|lll}
			\hline
			Model     & RG-1 & RG-2 & RG-L \\ \hline
			ABS$^\dag$       & 29.55$^*$  & 11.32$^*$  & 26.42$^*$  \\
			ABS+$^\dag$      & 29.78$^*$  & 11.89$^*$  & 26.97$^*$  \\
			Feats2s$^\dag$   & 32.67$^*$  & 15.59$^*$   & 30.64$^*$  \\
			RAS-Elman$^\dag$ & 33.78$^*$  & 15.97$^*$   & 31.15$^*$  \\
			Luong-NMT$^\dag$ & 33.10$^*$  & 14.45$^*$  & 30.71$^*$  \\
			s2s+att   & 34.23$^*$  & 15.52$^*$  & 31.57$^*$  \\ \hline
			FTSum$_c$     & 35.73$^*$   & 16.02$^*$   & 34.13   \\
			FTSum$_g$     & \textbf{37.27}   & \textbf{17.65}   & \textbf{34.24}   \\ \hline
		\end{tabular}
		\caption{ROUGE F1 performance. ``$^*$'' indicates statistical significance of the corresponding model with respect to the baseline model on the 95\% confidence interval in the official ROUGE script. RG refers to ROUGE for short.}
		\label{tb:rouge}
	\end{table}
	
	\subsection{Faithfulness Evaluation}
	Next, we conduct manual evaluation to inspect the faithfulness of the generated summaries.
	Specially, we randomly select 100 sentences from the test set.
	Then, we classify the generated summaries as FAITHFUL, FAKE or UNCLEAR.
	For the sake of a complete comparison, we present the results of our system FTSum$_g$ together with the the attentional s2s model s2s+att.
	As shown in Table~\ref{tb:validation}, about 30\% of the s2s-att outputs gives disinformation.
	This number greatly reduces to 6\% by our model.
	Nearly 90\% of summaries generated by our model is faithful, which makes our model far more practical.
	We find that s2s-att tends to copy the words closer to the predicate and regard them as its subject and object.
	However, this is not always reasonable and thus it is actually counterfeiting messages. 
	In comparison, the fact descriptions indeed designate the relations between a predicate and its subject and object.
	As a result, generation in line with the fact descriptions is usually able to keep the faithfulness.
	
	We illustrate the examples of defective outputs in Table~\ref{tb:faithfulness_examples}.
	As shown, att-s2s often attempts to fuse different parts in the source sentence to form the summary, no matter whether these phrases are relevant or not.
	For instance, att-s2s treats ``bosnian moslems'' as the subject of ``postponed'' and ``bosnia'' as the object of ``pulled out of'' in Example 1.
	By contract, since the fact description point out the actual subject and object, the output of our model is faithful.
	In fact, it is exactly the same as the target summary. 
	In Example 2, neither att-s2s nor our model achieves satisfactory performance.
	att-s2s again mismatches the object while our model fails to produce a complete sentence.
	To take a closer look, we find the target summary of this sentence is somewhat strange -- it merely focuses on the prepositional phrase (after taking a \#\# stoke...), rather than the main clause as usual.
	Since the main clause is hard to summarize and there is no high-quality fact description extracted, our model fails to give a complete summary.
	
	It is also noteworthy that, given multiple long fact descriptions, the generation of our model sometimes traps into one item.
	For instance, there are two long fact descriptions in Example 3 and our model only utilizes the first one for generation.
	As a result, despite the high faithfulness, the informativeness is somewhat damaged.
	Therefore, it seems more reliable to introduce the coverage mechanism~\cite{see2017get} to handle the cases like this one.
	We leave it as our future work.
	
	\begin{table}
		\centering
		\begin{tabular}{l|ll}
			\hline
			Model                    & Category    & Count \\ \hline
			\multirow{3}{*}{att-s2s} & FAITHFUL       & 68    \\
			& FAKE & 27    \\
			& UNCLEAR  & 5     \\ \hline
			\multirow{3}{*}{FTSum$_g$}   & FAITHFUL       & 87    \\
			& FAKE & 6     \\
			& UNCLEAR  & 7     \\ \hline
		\end{tabular}
		\caption{Faithfulness performance on the test set.}
		\label{tb:validation}
	\end{table}

	\begin{table*}[ht]
		\centering
		\begin{tabularx}{\linewidth}{p{1.2cm}|X}
			\hline
			\multicolumn{2}{c}{Example 1}                                                                                                                                                                                              \\ \hline
			\multicolumn{1}{l|}{Source}    & the repatriation of at least \#,\#\#\# bosnian moslems was postponed friday after the unhcr pulled out of the first joint scheme to return refugees to their homes in northwest bosnia .                      \\ \hline
			\multicolumn{1}{l|}{Relations} & unhcr pulled out of first joint scheme $|||$ repatriation was postponed friday $|||$ unhcr return refugees to their homes                                                                            \\ \hline
			\multicolumn{1}{l|}{Target}    & repatriation of bosnian moslems postponed                                                                                                                              \\ \hline
			\multicolumn{1}{l|}{att-s2s}   & (FAKE) \textbf{bosnian moslems} postponed after unhcr pulled out of \textbf{bosnia}                                                                                                                \\ \hline
			\multicolumn{1}{l|}{FTSum}     & (FAITHFUL) repatriation of bosnian moslems postponed                                                                                                               \\ \hline
			\multicolumn{2}{c}{Example 2}                                                                                                                                                                                              \\ \hline
			\multicolumn{1}{l|}{Source}    & davis love said he was thinking of making the world cup of golf a full time occupation after taking a \#\# stroke lead over japan in the event with us partner fred couples here on saturday .                                                    \\ \hline
			\multicolumn{1}{l|}{Relations} & making world cup full time occupation $|||$ taking \#\# stroke lead                                                                                                                       \\ \hline
			\multicolumn{1}{l|}{Target}    & americans lead UNK by \#\# strokes                                                                                                                                                        \\ \hline
			\multicolumn{1}{l|}{att-s2s}   & (FAKE) davis love says he is thinking of \textbf{the world cup }                                                                                                            \\ \hline
			\multicolumn{1}{l|}{FTSum}     & (UNCLEAR) love \textbf{in} the world cup of golf                                                                                                                            \\ \hline
			\multicolumn{2}{c}{Example 3}                                                                                                                                                                                              \\ \hline
			\multicolumn{1}{l|}{Source}    & the us space shuttle atlantis separated from the orbiting russian mir space station early saturday , after three days of test runs for life in a future space facility , nasa announced . \\ \hline
			\multicolumn{1}{l|}{Relations} & us space shuttle atlantis separated from orbiting russian mir space station $|||$ us space shuttle atlantis runs after three days of test for line in future space facility               \\ \hline
			\multicolumn{1}{l|}{Target}    & atlantis mir part ways after three-day space collaboration by emmanuel UNK                                                                                                                \\ \hline
			\multicolumn{1}{l|}{att-s2s}   & (UNCLEAR) space shuttle atlantis separated after \textbf{\#} days of test runs for life                                                                                     \\ \hline
			\multicolumn{1}{l|}{FTSum}     & (FAITHFUL) space shuttle atlantis separated from mir                                                                                                                          \\ \hline
		\end{tabularx}
		\caption{Examples of defective outputs. We use bold font to indicate the problematic parts.}
		\label{tb:faithfulness_examples}
	\end{table*}

	\subsection{Gate Analysis}
\begin{figure}
		\centering
		\includegraphics[width=0.7\linewidth]{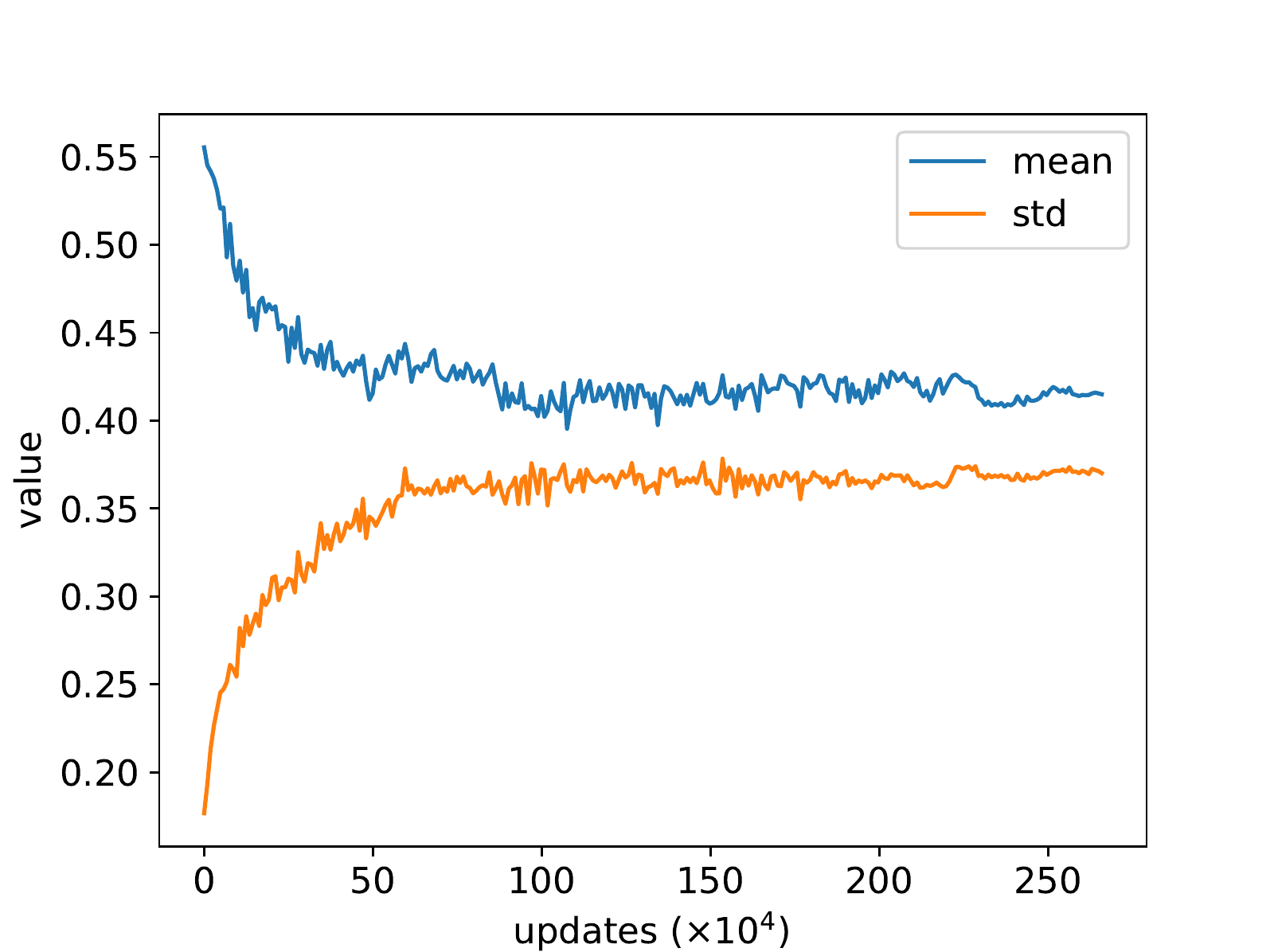}
		\caption{Gates change during training.}
		\label{fig:gate_change}
	\end{figure}

	As shown in Table~\ref{tb:rouge}, FTSum$_g$ achieves much higher ROUGE scores than FTSum$_c$.
	Now, we investigate what the gate network (Eq.~\ref{eq:context_gate}) actually learns.
	The changes of the gate values on the development set during training are shown in Fig.~\ref{fig:gate_change}.
	At the beginning, the average gate value exceeds 0.5, which means the generation is biased to the source sentence.
	As training proceeds, the model realizes that the fact descriptions are more reliable, resulting in a consecutive drop of the gate value.
	Finally, the average gate value is gradually stabilized to 0.415.
	Interestingly, the ratio of sentence and relation gate values i.e.,
	$(1 - 0.415)/0.415 \approx 1.41$, is extremely close to the ratio of copying proportions shown in Table~\ref{tb:relation_info} i.e.,
	$0.17/0.12 \approx 1.42$.
	It seems that our model predicts the copy proportion and normalizes it as the gate value.
	Then, look at the standard deviation of gates.
	To our surprise, its change is nearly anti-symmetric to the mean value.
	The final standard deviation reaches about 90\% of the mean gate value.
	Thus, still many sentences can dominate the generation.
	This strange observation urges us to carefully check the summaries with top/bottom-100 gate values in the development set.
	We find 10 fact descriptions in the top-100 cases are empty, and nearly 60\% contains the UNK tag.
	Our model believes these fact descriptions have not much worth to guide generation.
	Instead, there is no empty fact descriptions and only 1 UNK tag in the bottom 100 cases.
	Hence these fact descriptions are usually informative enough.
	In addition, we find the instances with the lowest gate values often hold the following (target summary; fact description) pair: 
	\begin{description}
		\item[Target] COUNTRY share prices close/open \#.\# percent higher/lower
		\item[Fact] COUNTRY share prices slumped/dropped/rose \#.\# percent
	\end{description}
	The extracted fact description itself is already a proper summary.
	That is why fact descriptions are particularly preferred in generation.
	
	\section{Related Work}
	Abstractive sentence summarization~\cite{chopra2016abstractive} aims to produce a shorter version of a given sentence while preserving its meaning.
	Unlike document-level summarization, it is impossible for this task to apply the common extractive techniques (e.g., \cite{cao2015ranking,cao2015learning}).
	Early studies for sentence summarization included rule-based methods~\cite{zajic2007multi}, syntactic tree pruning~\cite{knight2002summarization} and statistical machine translation techniques~\cite{banko2000headline}. 
	
	Recently, the application of encoder-decoder structures has attracted growing attention in this area.
	\cite{rush2015neural} proposed the ABS model which consisted of an attentive Convolutional Neural Network (CNN) encoder and an neural network language model decoder.
	\cite{chopra2016abstractive} extended their work by replacing the decoder with Recurrent Neural Network (RNN).
	\cite{nallapati2016abstractive} followed this line and developed a full RNN based sequence-to-sequence (s2s) framework~\cite{sutskever2014sequence}.
	Experiments on the Gigaword test set~\cite{rush2015neural} show that the above models achieve state-of-the-art performance.
	
	In addition to the direct application of the general s2s framework, researchers attempted to import various properties of summarization.
	For example, \cite{nallapati2016abstractive} enriched the encoder with hand-crafted features such as named entities and POS tags.
	These features played important roles in traditional feature based summarization systems.
	\cite{gu2016incorporating} found that a large proportion of words in the summary were copied from the source text.
	Therefore, they proposed CopyNet which considered the copying mechanism during generation.
	Later, \cite{ziqiang2017rnn} extended this work by directly measuring the copying mechanism within neural attentions.
	Meanwhile, they modified the decoder to reflect the rewriting behavior in summarization.
	Recently, \cite{see2017get} used the coverage mechanism to discourage repetition.
	There were also studies to modify the loss function to fit the evaluation metrics.
	For instance, \cite{ayana2016neural} applied Minimum Risk Training strategy to maximize the ROUGE scores of generated summaries.
	\cite{paulus2017deep} used reinforcement learning algorithm to optimize a mixed objective function of likelihood and ROUGE scores.
	
	Notably, previous researches usually focused on the improvement of summary informativeness.
	To the best of our knowledge, we are the first to explore the faithfulness problem of abstractive summarization.

	\section{Conclusion and Future Work}
	This paper investigates the faithfulness problem in abstractive summarization.
	We employ popular OpenIE and dependency parse tools to extract fact descriptions in the source sentence.
	Then, we propose the dual-attention s2s framework to force the generation conditioned on both source sentence and the fact descriptions.
	Experiments on the Gigaword benchmark demonstrate that our model greatly reduce fake summaries by 80\%.
	In addition, since the fact descriptions often condense the meaning of the sentence, the import of them also brings significant improvement on informativeness.
	
	We believe our work can be extended in various aspects.
	On the one hand, we plan to improve our decoder with the copying mechanism and coverage mechanism, which is further adapted to summarization.
	On the other hand, we are interested in the automatic evaluation of summary faithfulness.
	
	\section{ Acknowledgments}
	The work described in this paper was supported by Research Grants Council of Hong Kong (PolyU 152036/17E), National Natural Science Foundation of China (61672445 and 61572049) and The Hong Kong Polytechnic University (G-YBP6, 4-BCDV).

	\bibliographystyle{aaai}
	
	\bibliography{ijcai2016}
	
\end{document}